%% file: DCTSCD14.tex
\newif\ifboyscout                         %%
\boyscouttrue %% commented, WWW/boyscouts %%
\newif\ifpreparepdf                       %%
\preparepdftrue % hyperlinked pdf default %%
\newif\ifhighlightedits
\highlighteditstrue % used in resubmissions
\boyscoutfalse                 % public, hyperlinked
\ifboyscout
\documentclass[prl,aps,preprint,showpacs,superscriptaddress]{revtex4-1} %or {revtex4}
\else
\documentclass[prl,aps,twocolumn,showpacs,superscriptaddress]{revtex4-1} %or {revtex4}
\fi

\input{upodimDefs}

\begin{document}

        \title{
Estimating dimension of inertial manifold from unstable periodic orbits
        }

\author{X. Ding}
\affiliation{
                Center for Nonlinear Science, School of Physics,
                Georgia Institute of Technology,
                Atlanta, GA 30332-0430, USA
               }
\author{H. Chat\'e}
\affiliation{
Service de Physique de l'Etat Condens\'e, CEA, CNRS,
Universit\'e Paris-Saclay, 91191 Gif-sur-Yvette, France}%
\affiliation{
Beijing Computational Science Research Center, Beijing 100094, China}%
\author{P. Cvitanovi\'c}
\affiliation{
                Center for Nonlinear Science, School of Physics,
                Georgia Institute of Technology,
                Atlanta, GA 30332-0430, USA
               }
\author{E. Siminos}
\affiliation{Department of Physics, Chalmers University of Technology,
             Gothenburg, Sweden}
\author{K. A. Takeuchi}
%\email{kat@kaztake.org}
\affiliation{Department of Physics, Tokyo Institute of Technology,
             2-12-1 Ookayama, Meguro-ku, Tokyo 152-8551, Japan}%

\date{\today}

\begin{abstract}
We provide numerical evidence that a finite-dimensional inertial manifold
on which the dynamics of a chaotic dissipative dynamical system lives can
be constructed solely from the knowledge of a set of unstable
periodic orbits. In particular, we determine the dimension of the
inertial manifold for Kuramoto-Sivashinsky system, and find it to be
equal to the `physical dimension' computed previously via the
hyperbolicity properties of covariant Lyapunov vectors.
\end{abstract}

%\pacs{} %Predrag 2016-03-09
%\keywords{}

\maketitle

Dynamics in chaotic dissipative systems is expected to land, after a
transient period of evolution, on a finite-dimensional object in
\statesp\ called the inertial
manifold\rf{constantin_integral_1989,infdymnon,temam90,Foias1988a,Robinson1995}.
This is true even for infinite-dimensional systems described by partial
differential equations, and offers hope that their asymptotic dynamics
may be represented by a finite set of ordinary differential equations, a
drastically simplified description. The existence of a finite-dimensional
inertial manifold has been established for systems such as the
Kuramoto-Sivashinsky, the complex Ginzburg-Landau, and some reaction-diffusion
systems\rf{infdymnon}. For the Navier-Stokes flows its
existence remains an open problem\rf{temam90},
but dynamical studies, such as the determination of sets of \po s embedded in
a turbulent flows\rf{GHCW07,WiShCv15}, strengthen the case for a geometrical
description of turbulence. However, while mathematical approaches
may provide rigorous bounds on dimensions of inertial manifolds, a
constructive description of inertial manifolds remains a challenge.

Recent progress towards this aim comes from the viewpoint of linearized
stability analysis of spatio-temporally chaotic flows\rf{YaTaGiChRa08,TaGiCh11}.
Specifically, numerical investigations of the dynamics of {\cLvs}, made
possible by the algorithm developed in
\refrefs{ginelli-2007-99,GiChLiPo12}, have revealed that the tangent space
of a generic spatially-extended dissipative system is split into two
hyperbolically decoupled subspaces: a finite-dimensional subspace of
``entangled'' or ``physical'' Lyapunov modes (referred to in what follows
as the ``physical manifold''), which is presumed to capture all long-time
dynamics, and the remaining infinity of transient (``isolated,''
``spurious'') Lyapunov modes.
Covariant Lyapunov vectors span the Oseledec
subspaces\rf{lyaos,EckmannRuelle1985} and thus indicate the intrinsic
directions of growth or contraction at every point on the
physical manifold.
The dynamics of the vectors that span the physical manifold is entangled,
with frequent tangencies between them.
The {\transient} modes, on the other hand, are damped so strongly by the
dissipation, that they are isolated - at no time do they couple by
tangencies to the {\entangled} modes that populate the physical manifold.
In \refrefs{YaTaGiChRa08,TaGiCh11} it was conjectured that the physical
manifold provides a local linear approximation to the inertial manifold
at any point on the attractor, and that the dimension of the inertial
manifold is given by the number of the {\entangled} Lyapunov modes. Further
support for this conjecture was provided by
\refref{YaRa11}, which verified that the vectors connecting pairs of
recurrent points --points on the chaotic trajectory far apart in time but
nearby each other in \statesp-- are confined within the local tangent
space of the physical manifold.

These simulations of long time chaotic trajectories have been
successful in establishing that the physical manifold, defined everywhere
along a chaotic trajectory by locally flat tangent space, captures the
finite dimensionality of the inertial manifold, but they do not tell us
much about how this inertial manifold is actually laid out in \statesp.

In this letter, we go one important step further and show that the
finite-dimensional physical manifold can be precisely embedded in its
infinite-dimensional \statesp, thus opening a path towards its explicit
construction. The key idea\rf{DasBuch} is to populate the inertial
manifold by an infinite hierarchy of unstable time-invariant solutions,
such as periodic orbits, an invariant skeleton which, together
with the local ``tiles'' obtained by linearization of the dynamics,
fleshes out the physical manifold. Chaos can then be viewed as a walk on
the inertial manifold, chaperoned by the nearby unstable solutions
embedded in the physical manifold. Such unstable \po s have already been
used to compute global averages over chaotic dynamics, also for
spatially-extended systems, such as the
\KS\rf{Christiansen97,lanCvit07,SCD07} and Navier-Stokes\rf{GHCW07}
spatiotemporally chaotic flows.

In principle there are infinitely many unstable orbits, and each of them
{possesses} infinitely many Floquet modes. While in the example that we
study here we do not have a detailed understanding of the organization of
\po s (their symbolic dynamics), we do have sufficiently many of them to
be able to show that one only needs to consider a finite number of
unstable orbits to tile the physical manifold to a reasonable accuracy.
We also show, for the first time, that each local tangent
tile spanned by the Floquet vectors of an unstable periodic orbit
splits into a set of {\entangled} Floquet modes  and the remaining set of
{\transient} modes.
Furthermore, we verify numerically that the {\entangled} Floquet manifold
coincides locally with the physical manifold determined by the covariant
Lyapunov vectors approach.

Throughout this letter, we focus on the one-dimensional
\KSe\rf{KurTsu76,siv}, chosen here as a prototypical dissipative
partial differential equation that
exhibits spatiotemporal chaos\rf{cross93,Holmes96},
\begin{equation}
  u_t+u_x u+u_{xx}+u_{xxxx}=0, \quad x\in [0,L],
  \label{eq:ks}
\end{equation}
with a real-valued `velocity' field $u(x,t)$, and the periodic boundary
condition $u(x,t)=u(x+L,t)$. Following \refref{SCD07}, we fix the size at
$L=22$, which is small enough so that unstable orbits are still
relatively easy to determine numerically, and large enough for the \KSe\
to exhibit essential features of spatiotemporal chaos\rf{HNks86}.
Dynamical evolution traces out a trajectory in the $\infty$-dimensional
\statesp, $\op(t)=f^t(\op(0)$), with $\op(t)\equiv{}u(x,t)$, where the
time-forward map $f^t$ is obtained by integrating
\(\dot{\op} = \mathrm{\mathbf{\vel}}(\op)\) up to
time $t$. The linear stability of the trajectory is described by the
Jacobian matrix $\jMps^{t}(\op(0)) =\partial\op(t)/\partial \op(0)$,
obtained by integrating
\(\dot{\jMps} = \Mvar\,\jMps\),
where $\Mvar$ is the \stabmat\
\(\Mvar(\op) = \partial \mathrm{\mathbf{\vel}}(\op)/\partial \op\) .
We integrate the system \refeq{eq:ks} numerically, by a pseudo-spectral
truncation\rf{cox02jcomp,ks05com}  of
\(
 u(x,t)=\sum_{k=-\infty}^{+\infty} \ssp_k(t) e^{ i q_k x}
\,,\;   q_k = 2\pi\,k/L
\)
to a finite number of Fourier modes.
For the numerical accuracy required here we found $31$ Fourier modes
(62-dimensional \statesp) sufficient.
All orbits used here are found by a multiple shooting method and the
Levenberg-Marquardt algorithm (see \refref{SCD07} for details). A high
accuracy computation of {\em all} Floquet exponents and vectors for this
finite-dimensional \statesp\ (the key to all numerics presented here) has
been made possible by the algorithm recently developed in
\refref{DingCvit14}.
In our analysis, we use 200 \ppo s $\overline{ppo}_{\period{p}}$ and 200
\rpo s $\overline{rpo}_{\period{p}}$, labelled by their periods
$\period{p}$. These are the shortest period orbits taken from the set of
over 60\,000 determined in \refref{SCD07} by near-recurrence searches.
The method preferentially finds orbits embedded in the long-time
attracting set, but offers no guarantee that all orbits up to a given
period have been found.

The system is invariant under the Galilean transformations
$u(x,t)\to{}u(x-ct,t)+c$,
reflection $u(x,t)\to-u(-x,t)\equiv\sigma\,u(x,t)$,
and spatial translations
$u(x,t)\to{}u(x+\ell,t)\equiv{}g(\theta)\,u(x,t)$,
where $\theta=2\pi\ell/L$,
and ${\sigma}$ is the reflection operator.
The Galilean symmetry is reduced by setting the mean velocity
$\int\!dx\,u(x,t)$, a conserved quantity, to zero.
Due to the \On{2} equivariance of \eqref{eq:ks}, this system can have two
types of `relative' recurrent orbits
(referred to collectively as ``orbits'' in what follows):
\ppo s $u(x,0)={\sigma}u(x,\period{p})$
and
\rpo s $u(x,0)=g(\theta_{p})\,u(x,\period{p})$,
where
$g(\theta_{p})$ is the spatial
translation by distance $\shift_p=L\theta_{p}/2\pi$.
They are fixed points of maps
${\sigma}f^{\period{p}}$ and $g(\theta_{p})f^{\period{p}}$, respectively.
Their Floquet multipliers $\Lambda_j$ and vectors $\ve_j(\op)$ are the
eigenvalues and eigenvectors of Jacobian matrix $J_p =
{\sigma}J^{\period{p}}$ or $J_p = g(\theta_{p})J^{\period{p}}$ for
pre-periodic or \rpo s, respectively.
The Floquet exponents $\lambda_j$
(if complex, we shall only consider their real parts, with multiplicity 2)
are related to multipliers by $\lambda_j=\ln|\Lambda_j|/\period{p}$.
For an orbit $(\lambda_j,\ve_j)$ denotes the $j$th Floquet
(exponent, vector); for a chaotic trajectory it denotes the
$j$th Lyapunov (exponent, vector).

\begin{figure}[t!]
 %\centering
 \includegraphics[width=\hsize,clip]{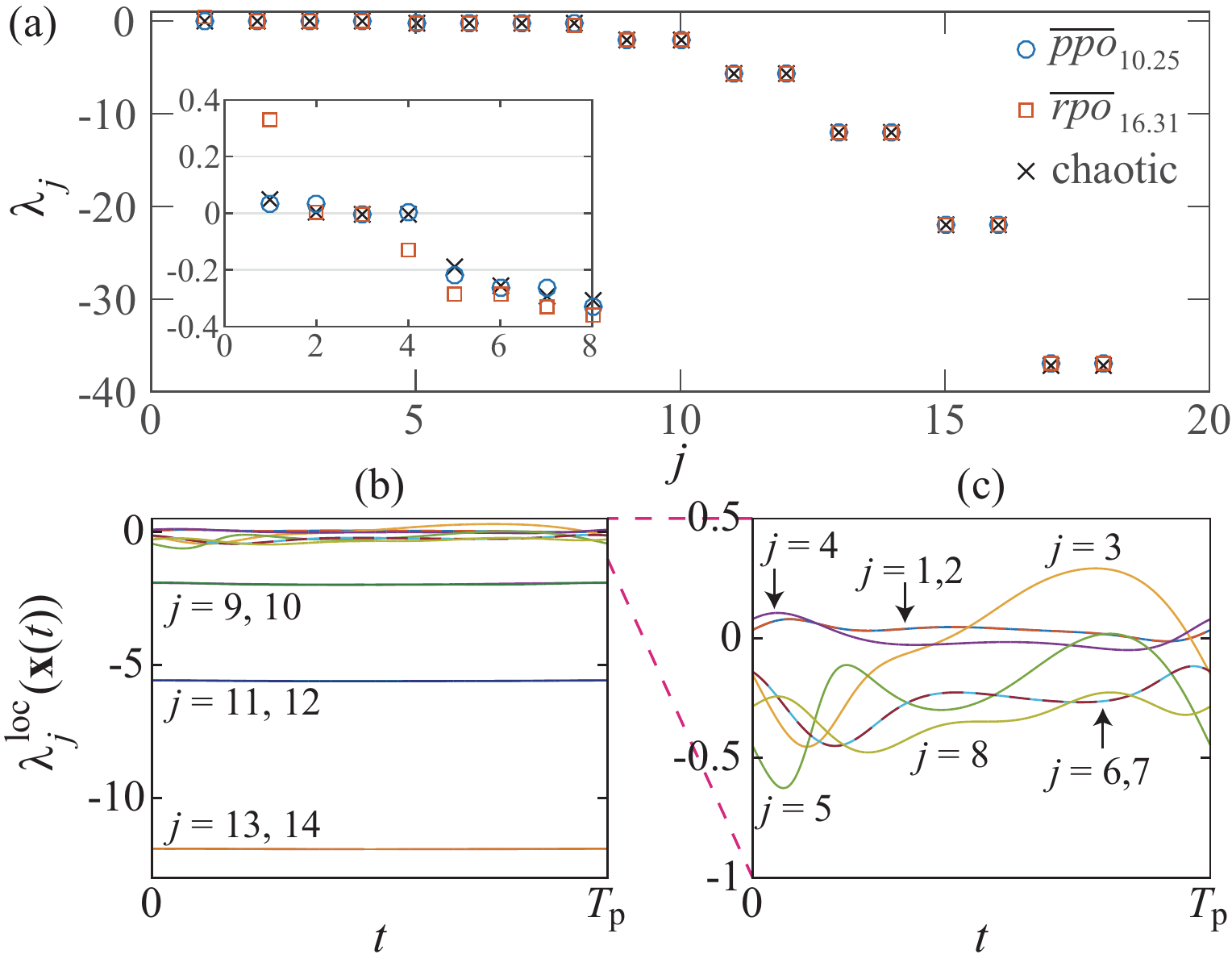}
 \caption{
(Color online)
(a) Floquet exponents for $\overline{ppo}_{10.25}$ (circles),
    $\overline{rpo}_{16.31}$ (squares), and Lyapunov exponents of a
    chaotic trajectory (crosses).
    The inset shows a close-up of the 8 leading exponents.
\PCedit{The number of the vanishing exponents is always two. Fourth
Lyapunov exponent is small but strictly negative,  $\lambda_4= -0.003$.}
(b) Time series of local Floquet exponents
    $\lambda_j(\op(t))$ for $\overline{ppo}_{10.25}$.
(c) Close-up of (b) showing the 8 leading exponents.
    Dashed lines indicate degenerate exponent pairs corresponding to
    complex Floquet multipliers.
 }
 \label{fig:ks22FloqExp}
\end{figure}

\refFig{fig:ks22FloqExp}\,(a) shows the Floquet exponents spectra for the two
shortest orbits, $\overline{ppo}_{10.25}$ and $\overline{rpo}_{16.31}$,
overlaid on the Lyapunov exponents computed from a chaotic trajectory.
The basic structure of this spectrum is shared by all 400
orbits used in our study\rf{KazzNt1}.
For chaotic trajectories, hyperbolicity between an arbitrary pair of
Lyapunov modes can be characterized by a property called the domination
of Oseledec splitting (DOS)\rf{PuShSt04,Bochi04}.
Consider a set of finite-time Lyapunov exponents
\begin{equation}
 \lambda_j^\tau(\op)
 \equiv
% \frac{1}{\tau}\ln \frac{||J^\tau(\op)\ve_j(\op)||}{||\ve_j(\op)||}
 \frac{1}{\tau}\ln ||J^\tau(\op)\ve_j(\op)||
\,,
\label{eq:ftle}
\end{equation}
with $L^2$ normalization $||\ve_j(\op)||=1$.
A pair of modes $j<\ell$ is said to fulfill `DOS strict ordering'
if $\lambda_{j}^\tau(\op) > \lambda_{\ell}^\tau(\op)$
along the entire chaotic trajectory, for $\tau$ larger than some lower
bound $\tau_0$. Then this pair is guaranteed not to have
tangencies\rf{PuShSt04,Bochi04}.
For chaotic trajectories, DOS turned out to be a useful tool to
distinguish {\entangled} modes from hyperbolically decoupled {\transient}
modes\rf{YaTaGiChRa08,TaGiCh11}.
\Po s are by definition the infinite-time orbits  ($\tau$ can be any repeat
of $\period{p}$), so generically all nondegenerate pairs of modes fulfill DOS.
Instead, we find it useful to define, by analogy to the `local Lyapunov
exponent'\rf{BosPos14}, the `local Floquet exponent' as the action of
the strain rate tensor\rf{Landau59a}
\(
2\,D(\op) = \transp{\Mvar(\op)}+\Mvar(\op)
\)
(where $\Mvar$ is the \stabmat) on the normalized
$j$th Floquet eigenvector,
\beq
\lambda_{j}(\op)
% = \transp{\ve_j(\op)} A(\op)\ve_j(\op),
% = \Re\left[\transp{\jEigvec[j](\ssp)}\Mvar(\ssp)\jEigvec[j](\ssp)\right]
= \transp{\ve_j(\op)}D(\op)\,\ve_j(\op)
= \lim_{\tau\rightarrow 0}\lambda_j^\tau(\op)
\,.
\ee{strainRateTens}
We find that time series of local Floquet exponents $\lambda_j(\op(t))$
indicate a decoupling of the leading
`\entangled' modes from the rest of the strictly ordered, strongly
negative exponents [\reffig{fig:ks22FloqExp}\,(b) and (c)].
Perhaps surprisingly, for every one of the 400 orbits we analyzed, the
number of the {\entangled} Floquet modes was \textit{always} 8, equal to the
previously reported number of the {\entangled} Lyapunov modes for this
system\rf{YaRa11,KazzNt1}.
This leads to our first surmise: (1) each individual orbit embedded in
the attracting set carries enough information to determine the  dimension
of the physical manifold.

\begin{figure}[tb]
\includegraphics[width=\hsize]{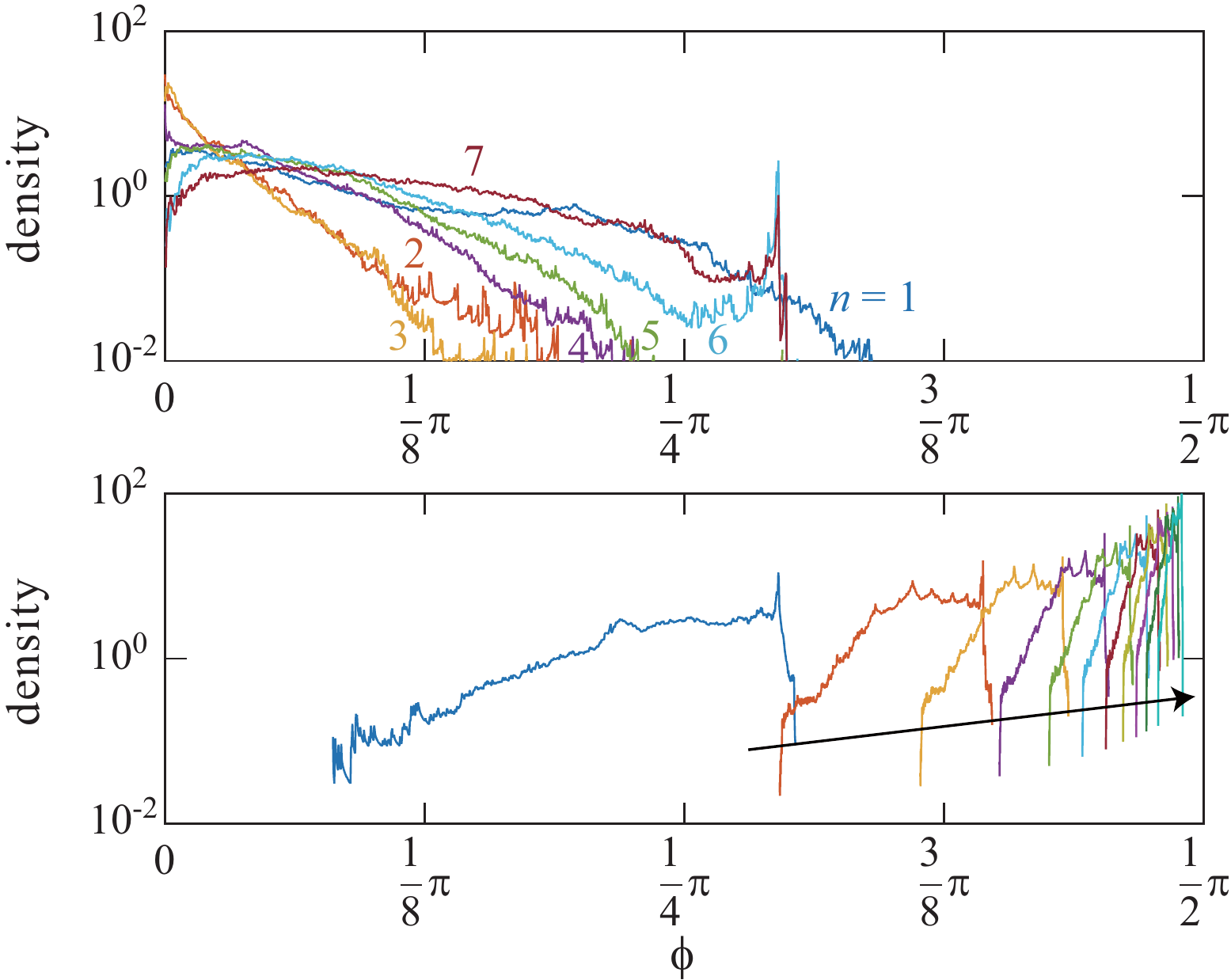}
 \caption{
(Color online) A histogram of the principal angles $\phi$ between $S_n$
(the subspace spanned by the $n$ leading Floquet vectors) and $\bar{S}_n$
(the subspace spanned by the remaining $d-n$ Floquet vectors),
accumulated over the 400 orbits used in our analysis. (top  panel) For
$n=1,2,\cdots,7$ ($S_n$ within the \entangled\ manifold) the angles can be
arbitrarily small. (bottom panel)
For $n=8,10,12,\cdots,28$ (in the order of the arrow),
for which all \entangled\ modes are contained in $S_n$,
 the angles are bounded away from unity.
 }
 \label{fig:ks22vecAngles}
\end{figure}

For an infinite-time chaotic trajectory, hyperbolicity can be assessed by
measuring the distribution of minimal principal
angles\rf{BjoGol73,Knyazev02} between any pair of subspaces spanned by
Lyapunov vectors\rf{ginelli-2007-99,YaTaGiChRa08,TaGiCh11}. Numerical
work indicates that as the {\entangled} and {\transient} modes are
hyperbolically decoupled, the distribution of the angles between these
subspaces is bounded away from zero, and that observation yields a sharp
{\entangled}-{\transient} threshold.

This strategy cannot be used for individual orbits, as each one is of a
finite period, and the minimal principal angle reached by a pair of
Floquet subspaces remains strictly positive.
Instead, we measure the angle distribution for a \textit{collection} of
orbits, and find that the {\entangled}-{\transient} threshold is as sharp
as for a long chaotic trajectory: \reffig{fig:ks22vecAngles} shows the principal angle
distribution between two subspaces $S_n$ and $\bar{S}_n$, with $S_n$
spanned by the leading $n$ Floquet vectors and $\bar{S}_n$ by the rest. As in the
{\cLvs} analysis of long chaotic trajectories\rf{YaTaGiChRa08}, the
distributions for small $n$ indicate strictly positive density as
$\phi\to0$. In contrast, the
distribution is strictly bounded away from zero angles for $n\geq 8$,
the number determined above by the local Floquet exponents analysis.
This leads to our second surmise:
(2) The distribution of principal angles for collections of \po s enables
us to identify a finite set of {\em {\entangled} Floquet modes}, the
analogue of the chaotic trajectories'  {\entangled} \cLv\ modes.

It is known, at least for low-dimensional chaotic attractors, that a
dense set of periodic orbits constitutes the skeleton of a strange
attractor\rf{DasBuch}. Chaotic trajectories meander around these orbits,
approaching them along their stable manifolds, and leaving them along their unstable
manifolds. If long-time trajectories are indeed confined to a
finite-dimensional physical manifold, such shadowing events should take
place within the subspace of {\entangled} Floquet modes of the
shadowed orbit. To analyze such shadowing, we
need to measure the distances between the chaotic trajectories and the
invariant orbits: the essential step here is {\em symmetry reduction},
\ie, replacement of a group orbit of states identical up to a symmetry
transformation by a single state. Since translation
$u(x,t)\to{}u(x+\ell,t)$ on a periodic domain implies a rotation
$\ssp_k(t)\to{}e^{iq_k\ell}\ssp_k(t)$ in Fourier space,
 we choose to send both
trajectories and orbits to the hyperplane $\im(\ssp_1)=0,\re(\ssp_1)>0$, called
the first Fourier-mode slice\rf{BudCvi14}, and measure the distances
therein. This transformation reads
\begin{equation}
  \label{eq:reduceSym}
  \hat{u}(x,t) = g(-\theta(t))u(x,t)
\end{equation}
 with $\theta(t)=\arg{}\ssp_1(t)$.
In the \slice, both \rpo s and \ppo s are reduced to \po s. From
\eqref{eq:reduceSym}, one easily finds how infinitesimal perturbations
$\delta{}u(x,t)$ are transformed\rf{KazzSuppl}.
This allows us to define the symmetry-reduced tangent space,
with the in-slice perturbations
$\delta\hat{u}(x,t)$, \JacobianM\ $\hat{J}^t(\sspRed)$,
Floquet matrix $\hat{J}_p(\sspRed)$
and Floquet vectors $\hat\ve_j(\sspRed)$.
The dimension of the \slice\ subspace is one less than the full \statesp:
\slice\ eliminates the marginal translational direction, while the
remaining Floquet multipliers $\Lambda_j$ are unchanged by the
transformation.
Therefore, for the system studied here, there are only seven {\entangled}
modes, with one marginal mode (time invariance) in the in-slice
description, instead of eight and two, respectively, in the full
\statesp\ description.
A shadowing of
an orbit $u_{p}(x,t')$ by a nearby chaotic trajectory $u(x,t)$ is
then characterized by the in-slice separation vector
\begin{equation}
  \label{eq:dif}
  \Delta \hat{u}(x,t) \equiv \hat{u}(x, t) -\hat{u}_{p}(x, t_{p}),
\end{equation}
where $t_{p}$ is chosen to minimize the in-slice distance
$||\Delta\hat{u}||$.

Now we test whether the separation vector $\Delta{}\hat{u}(x,t)$ is
confined to the tangent space spanned by the {\entangled} in-slice Floquet
vectors. To evaluate this confinement, one needs to take into account the
nonlinearity of the stable and unstable manifolds for finite
amplitude of $\Delta\hat{u}(x,t)$.
We decompose the separation vector as
\begin{equation}
 \Delta\hat{u}(x,t)=\hat{v}_n(x,t)+\hat{w}_n(x,t),  \label{eq:DiffVec}
\end{equation}
where $\hat{v}_n(x,t)$ is a vector in the subspace $\hat{S}_n$ spanned by
the leading $n$ in-slice Floquet vectors and $\hat{w}_k(x,t)$ is in
the orthogonal complement
of $\hat{S}_n$. If $n$ is large enough so that $\hat{S}_n$
contains the local approximation of the inertial manifold, we expect
$||\hat{w}_n||\sim||\hat{v}_n||^2\sim||\Delta\hat{u}||^2$ because of the
smoothness of the inertial manifold;
otherwise $||\hat{w}_n||$ does not vanish as $||\Delta\hat{u}||\to{}0$.
In terms of the angle $\varphi_n$ between $\hat{S}_n$ and
$\Delta\hat{u}$,
$\sin\varphi_n\sim||\hat{w}_n||/||\hat{v}_n||\sim||\Delta\hat{u}||$ for
$n$ above the threshold, while $\sin\varphi_n$ remains non-vanishing
otherwise.

\begin{figure}[h]
  \centering
 \includegraphics[width=\hsize,clip]{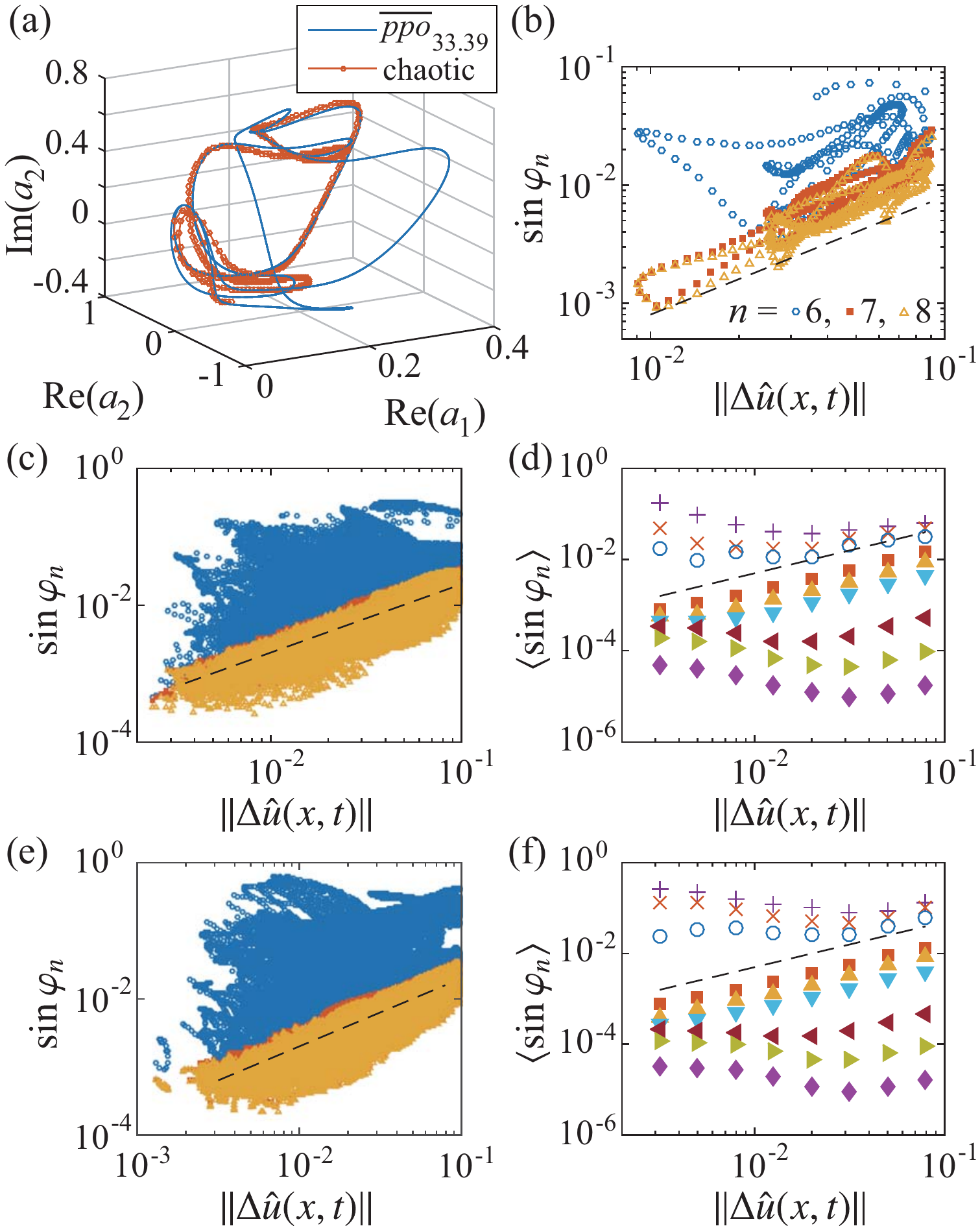}
  \caption{
(Color online)
(a) Shadowing event between a chaotic trajectory and $\overline{ppo}_{33.39}$,
drawn over $2\,\period{p}$.
(b) Parametric plot of $\sin\varphi_n(t)$ vs $||\Delta\hat{u}(x,t)||$
 during the single shadowing event shown in (a), for $n=6,7,8$.
(c) Same as (b), but a total of 230 shadowing events \ESedit{of $\overline{ppo}_{33.39}$} are used.
(d) Average of $\sin\varphi_n$ collected from the 230 shadowing events,
 taken within each bin of the abscissa,
 for $n=4,5,6,7,9,11,17,21,25$ from top to bottom.
(e)(f) Same as (c)(d), respectively,
 but for 217 shadowing events with $\overline{rpo}_{34.64}$.
The dashed lines show $\sin\varphi_n\propto||\Delta\hat{u}||$ in all panels.
 }
  \label{fig:ks22vecShadow}
\end{figure}

Following this strategy, we collected segments of a long-time chaotic
trajectory during which it stayed sufficiently close to a specific orbit
for at least one period of the orbit. \refFig{fig:ks22vecShadow}\,(a)
illustrates such shadowing event with respect to
$\overline{ppo}_{33.39}$. A parametric plot of $\sin\varphi_n(t)$ vs.
$||\Delta\hat{u}(x,t)||$ during this event is shown in
\reffig{fig:ks22vecShadow}\,(b) for $n=6,7,8$ (blue circles, red squares,
orange triangles, respectively). We can already speculate from such a
single shadowing event that $\sin\varphi_n$ does not necessarily decrease
with $||\Delta\hat{u}||$ for $n<7$, while it decreases linearly with
$||\Delta\hat{u}||$ for $n\geq7$. This threshold is clearly identified by
accumulating data for all the recorded shadowing events with
$\overline{ppo}_{33.39}$, \reffig{fig:ks22vecShadow}\,(c):
$\sin\varphi_n$ is confined below a line that depends linearly on
$||\Delta\hat{u}||$ if and only if $n\geq7$. We can see similarly clear
separation in the average of $\sin\varphi_n$ taken within each bin of the
abscissa [\reffig{fig:ks22vecShadow}\,(d)]. This indicates that for $n<7$
(empty symbols), typical shadowing events manifest significant deviation
of $\Delta\hat{u}$ from the subspace $\hat{S}_n$, whereas for $n\geq7$
(solid symbols) $\Delta\hat{u}$ is always confined to $\hat{S}_n$. We
therefore conclude that shadowing events are confined to the subspace
spanned by the leading 7 in-slice Floquet vectors, or equivalently, by
all the 8 {\entangled} Floquet vectors in the full \statesp. The
same conclusion was drawn for $\overline{rpo}_{34.64}$
[\reffig{fig:ks22vecShadow}\,(e) and (f)] and five other orbits
(not shown). We also verified that, when a chaotic trajectory approaches
an orbit, the subspace spanned by all {\entangled} Floquet modes of the
orbit coincides with that spanned by all {\entangled} Lyapunov modes of
the chaotic trajectory. This implies our third surmise: the {\entangled}
Floquet manifold coincides locally with the {\entangled} Lyapunov
manifold, with either capturing the local structure of the inertial
manifold.

In summary, we have used here the \KS\ system to demonstrate by six
independent calculations that the tangent space of a dissipative flow
splits into \entangled\ vs. \transient\ subspaces, and to determine the
dimension of its inertial manifold. The \emph{Lyapunov modes} approach of
\refrefs{ginelli-2007-99,YaTaGiChRa08,YaRa11,TaGiCh11,KazzNt1},
applied here to \KS\ on $L=22$ domain, identifies
    (1) the ``\entangled'' Lyapunov exponents, by the dynamics of
finite-time Lyapunov exponents, \eqref{eq:ftle}; and
    (2) the ``\entangled'' tangent manifold, or ``physical manifold,''
by measuring the distributions of angles between {\cLvs}.
The \emph{Floquet modes} approach\rf{DingCvit14} developed here shows that
    (3) Floquet exponents of each \emph{individual} orbit separate into
\entangled\ vs. \transient, \refFig{fig:ks22FloqExp};
    (4) for ensembles of orbits, the principal angles between hyperplanes
spanned by Floquet vectors separate the tangent space into \entangled\
vs. \transient, \reffig{fig:ks22vecAngles};
    (5) for a chaotic trajectory shadowing a given orbit the separation
vector lies within the orbit's Floquet \entangled\ manifold,
\reffig{fig:ks22vecShadow}; and
    (6) for a chaotic trajectory shadowing a given orbit the separation
vector lies within the chaotic trajectories \cLvs' \entangled\
manifold.

All six approaches yield the same inertial manifold dimension, reported
in earlier work\rf{YaRa11,KazzNt1}.
The Floquet modes / unstable periodic orbits approach is constructive, in
the sense that periodic points should enable us, in principle (but not
attempted in this letter), to tile the global inertial manifold by
local tangent spaces of an ensemble of such points.
Moreover, and somewhat surprisingly, our results on individual orbits'
Floquet exponents, \reffig{fig:ks22FloqExp}\,(b) and (c), and on
shadowing of chaotic trajectories, \reffig{fig:ks22vecShadow}, suggest
that \textit{each individual orbit} embedded in the attracting set
contains sufficient information to determine the
{\entangled}-{\transient} threshold.
However, the computation and organization of unstable periodic orbits is
still a major undertaking, and can currently be carried out only for
rather small computational domains\rf{SCD07,WiShCv15}.
The good news is that the {\entangled} Lyapunov modes
approach\rf{YaTaGiChRa08} suffices to determine the inertial manifold
dimension, as Lyapunov modes calculations only require averaging over
long chaotic trajectories, are much easier to implement, and can be
scaled up to much larger domain sizes than $L=22$ considered here.

We hope the computational tools introduced in this letter can eventually
contribute to solving outstanding issues of dynamical systems theory,
such as the existence of an inertial manifold in the transitional
turbulence regime of the \NSe, as well as to provide a practical guide to
numerically accurate truncations of infinite-dimensional systems
described by partial differential equations.

%%%%%%%%%%%%%%%%%%%%%%%%%%%%%%%%%%%%%%%%%%%%%%%%%%%%
\acknowledgements

We are indebted to
    \Private{
W. Hu,
Al Soyu,
N. Ott Gain,
and
    }
Ruslan\ L. Davidchack
for valuable input throughout this project,
in particular for the 60\,000 \rpo s that started it,
and to
Mohammad\ M. Farazmand
for
a critical reading of the manuscript.
X.D. and P.C.\ were supported by NSF~DMS-1211827.
P.C.\ thanks the family of late G.~Robinson,~Jr.\ for continued support.
K.A.T. acknowledges support by KAKENHI (No. 25707033 from JSPS and No.
25103004 `Fluctuation \& Structure' from MEXT in Japan) and the JSPS
Core-to-Core Program `Non-equilibrium dynamics of soft matter and
information'.

%%%%%%%%%%%%%%%%%%%%%%%%%%%%%%%%%%%%%%%%%%%%%%%%%%%%

\bibliography{DCTSCD14}

%%%%%%%%%%%%%%%%%%%%%%%%%%%%%%%%%%%%%%%%%%%%%%%%%%%%
        \Private{\newpage
\input{../lyapunov/DCTSCD14}
        }
%%%%%%%%%%%%%%%%%%%%%%%%%%%%%%%%%%%%%%%%%%%%%%%%%%%%

\end{document}

%% file: upodimDefs.tex
    \usepackage{graphicx,amsmath,amssymb,amsbsy,bm}
    \usepackage{amsfonts,latexsym}
    \usepackage{ifthen}
    \usepackage[percent]{overpic}

\ifpreparepdf
    \usepackage{color}
    \usepackage[colorlinks]{hyperref} %% hyperlinks
\else % prepare B&W file
    \usepackage{hyperref}
\fi

    \graphicspath{{../xiong/figures/}} %{../figs/}{../Fig/}}  %% directories with graphics files

\ifpreparepdf % hyperlinked pdf, keep homepage flexible:

    \newcommand{\arXiv}[1]{
              {\tt \href{http://arXiv.org/abs/#1}{arXiv:#1}}}
\else  %% prepare for postscript printing:

    \newcommand{\arXiv}[1]{ {\tt arXiv:#1}}
\fi

\ifhighlightedits
    
\else
    
\fi

        \ifboyscout % if draft, display comments in text
   \newcommand{\PCedit}[1]{{\color{blue}#1}}
   \newcommand{\PC}[2]{\begin{quote}\PCedit{[#1 Predrag] #2}\end{quote}}
   \newcommand{\RLDedit}[1]{{\color{red}#1}}
   \newcommand{\RLD}[2]{\begin{quote}\RLDedit{[#1 Ruslan] #2}\end{quote}}
   \newcommand{\Xiongedit}[1]{{\color{green}#1}}
   \newcommand{\Xiong}[2]{\begin{quote}\Xiongedit{[#1 Xiong] #2}\end{quote}}
   \newcommand{\ESedit}[1]{{\color{magenta}#1}}
   \newcommand{\ES}[2]{\begin{quote}\ESedit{[#1 Evangelos] #2}\end{quote}}
   \newcommand{\Private}[1]{{\color{blue}#1}}
   \newcommand{\KTedit}[1]{{\color{cyan}#1}}
   \newcommand{\KT}[2]{\begin{quote}\KTedit{[#1 Kazz] #2}\end{quote}}
   \newcommand{\HCedit}[1]{{\color{red}#1}}
   \newcommand{\HC}[2]{\begin{quote}\HCedit{[#1 Hugues] #2}\end{quote}}
        \else  % drop comments
   \newcommand{\PC}[2]{}{}
   \newcommand{\PCedit}[1]{#1}
   \newcommand{\RLD}[2]{}{}
   \newcommand{\RLDedit}[1]{#1}
   \newcommand{\Xiong}[2]{}{}
   \newcommand{\Xiongedit}[1]{#1}
   \newcommand{\ES}[2]{}{}
   \newcommand{\ESedit}[1]{#1}
   \newcommand{\HC}[2]{}{}
   \newcommand{\HCedit}[1]{#1}
   \newcommand{\Private}[1]{}
   \newcommand{\KT}[2]{}{}
   \newcommand{\KTedit}[1]{#1}
        \fi

\newcommand{\rf}     [1] {~\cite{#1}}
\newcommand{\refref} [1] {ref.~\cite{#1}}

\newcommand{\refrefs}[1] {refs.~\cite{#1}}
\newcommand{\refRefs}[1] {Refs.~\cite{#1}}
\newcommand{\refeq}  [1] {(\ref{#1})}
\newcommand{\reffig} [1] {fig.~\ref{#1}}
\newcommand{\refFig} [1] {Fig.~\ref{#1}}

\renewcommand{\eqref}[1]{Eq.~(\ref{#1})}

\newcommand{\ve}{\bm{e}}
\newcommand{\op}{\mathrm{\mathbf{x}}}

\DeclareMathOperator{\im}{Im}
\DeclareMathOperator{\re}{Re}

\newcommand{\JacobianM}{Jacobian matrix} %
\newcommand{\stabmat}{stability matrix}     % stability matrix, velocity gradients
\newcommand{\cLv} {covariant Lyapunov vector} % Ginelli et al
\newcommand{\cLvs} {covariant Lyapunov vectors} % Ginelli et al
\newcommand{\transient}{transient} % ChaosBook.org
\newcommand{\entangled}{entangled}  % ChaosBook.org
\newcommand{\ppo}{pre-periodic orbit}

\newcommand{\beq}{\begin{equation}}

\newcommand{\eeq}{\end{equation}}
\newcommand{\ee}[1] {\label{#1} \end{equation}}
\newcommand{\bea}{\begin{eqnarray}}

\newcommand{\eea}{\end{eqnarray}}

\newcommand{\ie}{{i.e.}}            % APS

\newcommand{\statesp}{state space}

\newcommand{\Mvar}{\ensuremath{A}}  % stability matrix
\newcommand{\jMps}{\ensuremath{J}}   % jacobian matrix, phase space/state space

\newcommand\period[1]{{\ensuremath{T_{#1}}}}         %continuous cycle period
\newcommand{\po}{periodic orbit}
\newcommand{\Po}{Periodic orbit}
\newcommand{\rpo}{rela\-ti\-ve periodic orbit}
\newcommand{\bseq}{\begin{subequations}}
\newcommand{\eseq}{\end{subequations}}

\newcommand{\NSe}{Navier-Stokes equations}

\newcommand{\slice}{slice}

\newcommand{\sspRed}{\ensuremath{\hat{\ssp}}}    % reduced state space point Jan 2012
\newcommand{\On}[1]{\ensuremath{\textrm{O}(#1)}}
\newcommand{\KS}{Kuramoto-Siva\-shin\-sky}
\newcommand{\KSe}{Kuramoto-Siva\-shin\-sky equation}

\newcommand{\eigExp}[1][]{
     \ifthenelse{\equal{#1}{}}{\ensuremath{\lambda}}{\ensuremath{\lambda^{(#1)}}}}
\newcommand{\eigRe}[1][]{
     \ifthenelse{\equal{#1}{}}{\ensuremath{\mu}}{\ensuremath{\mu^{(#1)}}}}
\newcommand{\eigIm}[1][]{
     \ifthenelse{\equal{#1}{}}{\ensuremath{\omega}}{\ensuremath{\omega^{(#1)}}}}

\newcommand{\vel}{\ensuremath{v}}   % state space velocity
\newcommand{\ssp}{\ensuremath{a}}                % state space point
\newcommand{\shift}{\ell}
\newcommand{\transp}[1]{{#1}{}^\top}

%% file: DCTSCD14.bbl
\begin{thebibliography}{35}
\bibitem [{Kaz({\natexlab{a}})}]{KazzNt1}%
  \BibitemOpen
  \href@noop {} {} ({\natexlab{a}}),\ \bibinfo {note}
  {{\refRefs{YaTaGiChRa08,TaGiCh11,YaRa11} include the marginal Galilean
  symmetry mode in the mode count; here this mode is absent, as we have set
  $\int{}u(x,t)dx=0$. Consequently, the number of the {\entangled} modes (the
  dimension of the physical manifold) differs by one.}}\BibitemShut {Stop}%
\end{thebibliography}
